# Injection locking of optomechanical oscillators via acoustic waves

Ke Huang, and Mani Hossein-Zadeh

*Center for High Technology Materials (CHTM), University of New Mexico, 1313 Goddard SE, NM 87106, USA*

Injection locking is a powerful technique for synchronization of oscillator networks and controlling the phase and frequency of individual oscillators using similar or other types of oscillators. Here, we present the first demonstration of injection locking of a radiation-pressure driven optomechanical oscillator (OMO) via acoustic waves. As opposed to previously reported techniques (based on pump modulation or direct application of a modulated electrostatic force), injection locking of OMO via acoustic waves does not require optical power modulation or physical contact with the OMO and it can easily be implemented on various platforms. Using this approach we have locked the phase and frequency of two distinct modes of a microtoroidal silica OMO to a piezoelectric transducer (PZT). We have characterized the behavior of the injection locked OMO with three acoustic excitation configurations and showed that even without proper acoustic impedance matching the OMO can be locked to the PZT and tuned over 17 kHz with only -30 dBm of RF power fed to the PZT. The high efficiency, simplicity and scalability of the proposed approach paves the road toward a new class of photonic systems that rely on synchronization of several OMOs to a single or multiple RF oscillators with applications in optical communication, metrology and sensing. Beyond its practical applications, injection locking via acoustic waves can be used in fundamental studies in quantum optomechanics where thermal and optical isolation of the OMO are critical.

## 1. INTRODUCTION

Optomechanical oscillation is among a relatively new set of effects enabled by the coupling between the optical and mechanical modes of optomechanical resonators (resonators that can sustain coupled optical and mechanical modes) via the radiation pressure of the resonant optical field. Self-sustained optomechanical oscillation based on radiation pressure was first reported in silica microtoroids [1-5] and triggered the development of many other optomechanical resonators [6-9]. These devices comprise a new class of RF oscillators called optomechanical oscillators (OMOs) that are driven by radiation pressure without the need for any electronic component. Low power consumption, small size, low phase noise and all-optical operation, makes these oscillators potential candidates for replacing electronic oscillators in certain RF-photonic communication and sensing applications. All-optical RF down-conversion in optical domain [10,11] and mass sensing [12,13] are among the most important reported applications that have benefited from the unique properties of OMOs. For all applications, stability of OMO and control over its phase and frequency are not only critical for the performance of the system, but also enable new functionalities. The frequency of an OMO is determined by its mechanical eigenmodes and therefore the microresonator size and structure. Typically a single OMO can support few oscillation frequencies associated with mechanical modes that are strongly coupled to high quality (high-Q) optical modes of the cavity. These modes can be selected by adjusting the laser wavelength and coupling strength near optical resonant wavelengths with sufficient quality factor [4,10,14]. For an isolated OMO, fine tuning of each oscillation frequency over a limited range can be achieved by changing the optical power (through optical spring effect) as well as microresonator temperature [14,15]. Alternatively similar to other self-sustained oscillators, the oscillation frequency of OMO can be controlled by injection locking to another oscillator [16-19].

Injection locking that has been extensively studied in electronic [20,21] and photonic oscillators (lasers) [22], not only provides control over the oscillation frequency and phase, but also enables synchronization of multiple oscillators to each other or to an external source. In general injection locking involves coupling (injecting) a periodic signal with a frequency close to the oscillation frequency into the oscillator. If the amplitude of the coupled signal is large enough, the frequency and phase of the oscillator are pulled and locked to that of the signal and therefore to the signal source. Basically the injected signal generated by the "master" oscillator acts as a perturbation for the "slave" oscillator; so the physical nature of the injected signal should be similar to one of the oscillating parameters in the slave oscillator. As such in electronic oscillators the injected signal can be an oscillating voltage, current or magnetic field and in lasers the injected signal is a coherent optical wave. Similarly in an optomechanical oscillator the injected signal can be a modulated optical power (perturbing the circulating optical power), a periodic mechanical force, or a mechanical wave (perturbing the mechanical motion).

The first observation of injection locking of an OMO was reported based on modulated optical pump [16], where the amplitude of the optical pump was partially modulated using an electro-optic modulator. Basically a small portion of the input power that was modulated at a frequency near $f_{OMO}$, acted as the injection signal and the OMO was locked to the RF source that was driving the modulator. Later synchronization of multiple OMOs using this approach was theoretically analyzed [23] and experimentally demonstrated for two OMOs [18,19]. While feeding a modulated optical pump to multiple OMOs in parallel or series configuration seems to be a trivial solution for synchronizing multiple OMOs, the fact that all these OMOs should have the same exact resonant optical wavelength, makes its practical implementation a very challenging task (in particular for oscillator networks). As fabrication of high-Q optomechanical cavities with the same exact resonant optical wavelengths is nearly impossible, these experiments require active thermal tuning of the corresponding optical cavities and therefore individual electrical contact with each OMO.

Recently it has been shown that OMO can be locked to an RF drive using electro-mechanical force directly applied on the OMO [24,25]. In this approach a metallic electrode that is deposited on top of the OMO (in this case a toroidal silica microcavity) is used to convert the RF voltage to a modulated force. This approach suffers from several shortcomings: 1) deposition of metallic electrode on the optomechanical resonators not only makes the fabrication process complicated, but also degrades the optical and mechanical quality factor (this may explain the large threshold pump power in Ref. 25 that is more than one order of magnitude larger than similar OMOs).

Clearly this problem is much more serious for nano scale optomechanical resonators such as zipper microcavities [7,9], small microdisks [6] or spoke supported microrings [8] (as their optical and mechanical quality factors are extremely sensitive to any perturbation). 2) In order to drive the electrodes, each OMO should be electrically connected to the RF source. While in a lab setting and for a single device the signal can be applied using special RF microprobes, in an integrated system the electric connection is a major challenge and limits the scalability of this technique.

In the present work, for the first time we report injection locking of OMO via acoustic waves excited by an external electromechanical oscillator. We demonstrate that if the acoustic waves can stimulate the mechanical mode coupled to the optical resonance, they can also pull and lock the frequency and phase of the optomechanical oscillation to the frequency and phase of the external oscillator that generates them. We show injection locking can occur with an acoustic excitation that generates a mechanical amplitude modulation as small as 5% of the original optomechanically generated modulation. As long as the acoustic waves reaching the OMO can generate sufficient mechanical vibration, the electro-mechanical transducer can be attached or fabricated at any location on the carrier chip without affecting the optomechanical properties of the OMO and interfering with its operation.

As such injection locking of OMOs via acoustic waves is superior to the previously reported techniques since by eliminating the need for physical contact with the microresonator and modulation of the optical pump power, it opens a wide range of possibilities for injection locking and synchronization of multiple OMOs using an external oscillator at reduced cost and complexity. Low power and large scale injection locking and synchronization of OMOs may benefit many applications such as optomechanical RF signal processing, optical communication and sensing.

Beyond its engineering applications, this new technique can be used in fundamental studies in quantum measurement, quantum optomechanics and nonlinear dynamics of coupled oscillators where physical isolation is critical and phase/frequency control should be achieved with minimal interference with OMO's intrinsic properties and optical feedback.

For the proof of concept demonstration of this approach, we used a toroidal silica microcavity as the optomechanical resonator/oscillator and a piezoelectric transducer as the electromechanical oscillator. While silica microtoroid has been selected because of its simplicity and relatively low threshold power and phase noise in atmospheric pressure [14], with proper design the same approach can be used to injection lock nearly any OMO (down to nanoscale) without the need to modify its structure.

We have observed and characterized the injection locking of two distinct mechanical modes of the selected microtoroidal OMO to an external piezoelectric transducer (PZT) via acoustic waves. The PZT was physically attached to three different positions on the chip that carried the OMO. Using a combination of finite element mechanical modeling for one of the modes and time-domain coupled differential equations, we verified that the behavior of the measured lock range as a function of RF input power fed to the PZT was in agreement with the classical theory of optomechanical oscillation. As such similar systems can be designed and optimized simply by finite element modeling of the acoustic energy exchange between the transducer and the selected mechanical mode and using the outcomes in the general coupled time-domain differential equations governing the optomechanical oscillation.

## 2. EXPERIMENT

Fig. 1(a) shows the experimental arrangement used for characterizing the microtoroidal OMO and demonstration of injection locking via acoustic waves. Optical power from a tunable laser ($\lambda_{laser}$~1550 nm) is coupled to high-$Q$ Whispering-Gallery modes (WGMs) circulating inside the microtoroidal optical cavity using a standard tapered silica fiber [1-3]. The coupling gap between the tapered fiber and the microtoroid is precisely controlled with a nanopositioner. A photodetector (bandwidth=150 MHz) is used to convert the optical power to electric signal for time and frequency domain analysis. Using the standard OMO characterization procedure [14] -- in the absence of acoustic excitation-- the high-Q WGMs with strongest coupling to two mechanical eigenmodes of the microtoroid were identified. Near each optical mode the coupling gap and wavelength detuning ($\Delta\lambda=\lambda_{laser}-\lambda_0$, $\lambda_0$: resonant wavelength of the corresponding optical mode) are optimized to obtain the minimum optomechanical threshold power ($P_{th}$) [1-5].

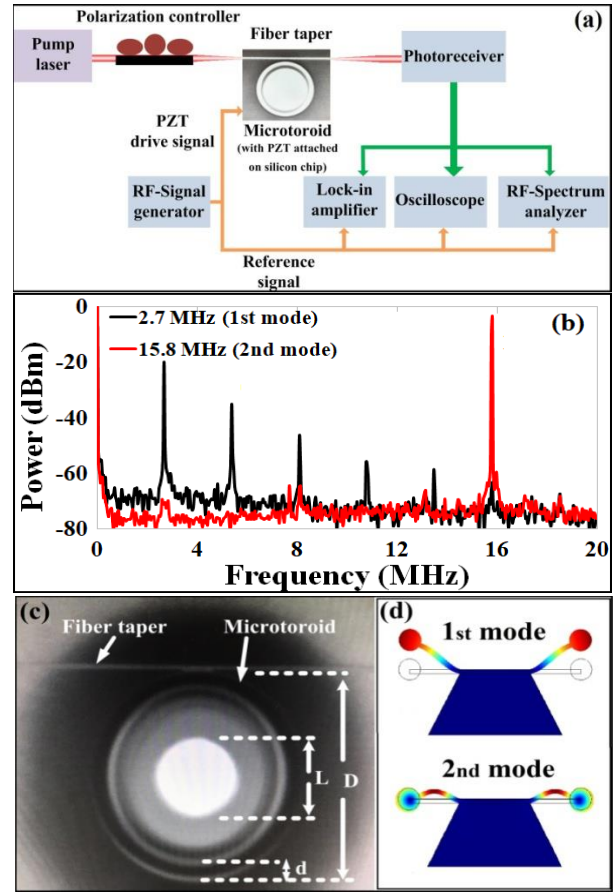

Fig. 1. (a) Experimental arrangement. (b) Measured RF spectrum of the transmitted optical power at $\Delta\lambda_1=-0.38\lambda_1/Q_{L1}$ (black), $\Delta\lambda_2=-0.42\lambda_2/Q_{L2}$ (red) corresponding to optomechanical oscillation of mode-1 with a frequency of 2.7 MHz and mode-2 with a frequency of 15.8 MHz. For mode-1 the higher harmonics (generated by the nonlinear optical transfer function of the cavity [26]) can be also observed. The measured mechanical quality factor ($Q_{mech}$) is 119 for the mode-1 and 360 for mode-2. (c) Top view micrograph of the silica microtoroid coupled to the silica fiber taper. The microtoroid has a major diameter of $D$=76 μm and minor diameter of $d$=9.7 μm. The diameter of the supporting silicon pillar is $L$=31.5 μm. (d) Mechanical deformation associated with mode-1 ($f_{OMO,1}$=2.7 MHz) and mode-2 ($f_{OMO,2}$=15.8 MHz) (calculated based on FEM using COMSOL software).

The first mechanical mode oscillates at $f_{OMO,1}$=2.7 MHz and is excited by an optical mode with a resonant wavelength of $\lambda_{01}$=1559.1 nm and loaded quality factor of $Q_{L1}$= 3.3×10$^6$. The second mechanical mode oscillates at $f_{OMO,2}$=15.8 MHz and is excited by an optical mode with a resonant wavelength of $\lambda_{02}$=1558.7 nm and loaded quality factor of $Q_{L2}$= 6.1×10$^6$. The measured threshold optical input power for exciting these oscillations were $P_{th,1}$=90 μW and $P_{th,2}$ =400 μW, respectively. Fig. 1(b) shows the RF spectrum of the transmitted optical power when $\lambda_{laser}$ was tuned near $\lambda_{01}$ (black trace) and $\lambda_{02}$ (red trace) while optical input power ($P_{o,in}$) was larger than $P_{th}$ for the corresponding mechanical mode. Fig. 1(c) shows the top-view micrograph of the silica microtoroid used in this experiment. After careful measurement of microtoroid dimensions, we used finite element modeling (COMSOL software) to identify the mechanical eigenmodes associated with the measured oscillation frequencies. Fig.1 (d) shows the calculated mechanical deformation associated with these modes indicating that mode-1 ($f_{OMO,1}$=2.7 MHz) is a flapping mode and mode-2 ($f_{OMO,2}$=15.8 MHz) is a breathing radial mode. Both modes are coupled to the circulating optical power through radial component of the microtoroid displacement ($\Delta R$). As such the modulation amplitude of the output optical signal is proportional to $\Delta R$ [14]. The calculated effective mass is 40 pg for the first mode and 36 pg for the second mode. Both modes were coupled to the optical mode with a coupling rate of 5.07 GHz/nm. Note that the ability of exciting and monitoring two distinct optomechanical modes of the OMO is important to demonstrate the possibility of injection locking of two different modes and showing the dependence of locking strength on mechanical deformation for a given acoustic excitation.

To study injection locking via acoustic waves, an external piezoelectric actuator is attached to the silicon chip that carries the OMO. The silicon chip has a dimension of 15 mm (L) × 4.5 mm (W) × 0.3 mm (H) and the ceramic actuator is a disk with a diameter of 20 mm and thickness of 0.2 mm. The selected piezo transducer (PZT) is designed to sustain mechanical oscillations through its thickness mode at a resonant frequency of 10.1 MHz. However by adjusting the drive frequency it can oscillate at a wide frequency range from 2 to 16 MHz with slightly lower efficiency and with a FWHM line-width smaller than ~80 Hz. We examined three configurations for exciting the mechanical modes of the microtoroid via acoustic waves generated by the PZT. These configurations are shown in Fig. 2(a)-(c): In configuration-1 (Fig. 2(a)), The PZT is attached to the bottom of the silicon chip right below the OMO, configuration-2 (Fig.2(b)) is similar to configuration-1 but the PZT is moved about 4 mm away from OMO; finally in configuration-3 (Fig.2(c)) the PZT is rotated 90 degrees and is attached to the side of the silicon chip 4 mm away from the OMO.

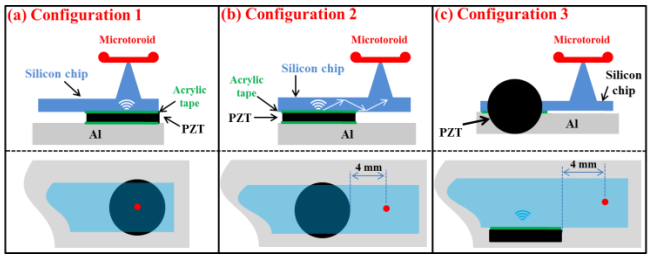

Fig. 2. Three configurations used to study the injection locking: (a) Configuration-1: The PZT is attached to the button of the silicon chip right below the microtoroid, (b) Configuration-2: The PZT is attached to the bottom of the silicon chip but 4 mm away from the microtoroid in the horizontal direction, (c) Configuration-3: The PZT is attached to the side edge of the silicon chip 4 mm away from the microtoroid.

In all configurations, the piezo transducer is attached to the silicon chip using an acrylic double-sided tape (thickness=70 μm) and is driven by a sinusoidal wave generated by an RF source. For this proof of concept demonstration the acoustic impedance of the PZT is not matched to that of the silicon chip as such a relatively small portion of the acoustic energy produced by the PZT is transferred to the silicon chip (only 27% and 15% of the acoustic energy generated by the PZT is transmitted to the silicon chip at 2.7 MHz and 15.8 MHz respectively). In principle using proper acoustic impedance matching layers between PZT and the chip 100% of the acoustic energy can be transferred to the silicon chip within the operational bandwidth of the PZT. For each optomechanical mode and configuration, the impact of the acoustic waves (generated by the PZT) on the OMO is evaluated by varying the power and frequency of the RF signal delivered to the PZT. The spectrum of the modulated output power and the relative phase between OMO and the RF signal are measured using an RF spectrum analyzer and a lock-in amplifier (as shown in Fig. 1(a)).

## 3. RESULTS

For the initial demonstration OMO injection locking was examined using configuration-2. Fig. 3(a) shows the RF spectrum of the modulated optical output power near $f_{OMO,1}$ in the presence (red trace) and absence (black trace) of acoustic excitation when $P_{o,in}$=2×$P_{th,1}$. Here the RF power delivered to the PZT ($P_{PZT}$) is –40 dBm and its frequency (=$f_{PZT}$) is 1.7 kHz smaller than $f_{OMO,1}$. Fig. 3 (b) shows the RF spectrum of the optical output power near $f_{OMO,2}$ in the presence (red trace) and absence (black trace) of acoustic excitation when $P_{o,in}$=1.4×$P_{th,1}$. Here the RF power delivered to the PZT ($P_{PZT}$) is –5 dBm and its frequency (=$f_{PZT}$) is 1.24 kHz larger than $f_{OMO,2}$. Note that not only a major portion of the acoustic energy is lost due to impedance mismatch and material loss, but also only a small fraction of the total energy delivered to the OMO couples to the desired mode. It is apparent that the injected acoustic wave pulls $f_{OMO,1}$ and $f_{OMO,2}$ and locks them to $f_{PZT}$. As expected the locking process also reduces the OMO line-width. Within the resolution bandwidth of the RF spectrum analyzer in 10 Hz, the oscillation liewidth of the OMO was equal to that of the RF oscillator.

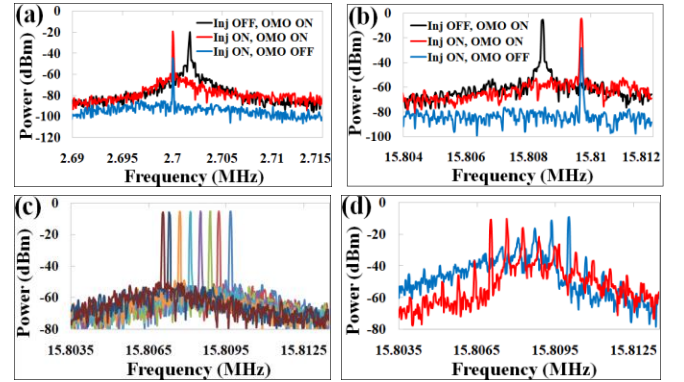

Fig. 3. Injection locking of mode-1 and mode-2 using configuration-2. (a) Measured spectrum of the optical power in the presence (red) and absence (black) of the injection signal (blue) for the 1st optomechanical mode. (b) Measured spectrum of the optical power in the presence (red) and absence (black) of the injection signal (blue) for the 2nd optomechanical mode. (c) Measured spectrum of the 2nd optomechanical mode tuned by the injection signal. (d) Measured spectrum of the 2nd optomechanical mode while the frequency of the injected signal is tuned slightly beyond the lock range. Note, in (a) $P_{PZT}$=-40 dBm, $P_{o,in}$=2.0×$P_{th}$, and $\eta$=0.138. In (b)-(d) $P_{PZT}$=-5 dBm, $P_{o,in}$=1.4×$P_{th}$, and $\eta$=0.058.

For both modes we have also measured optical modulation spectrum due excitation of the mechanical mode by the PZT (blue trace) by lowering $P_{o,in}$ to 0.8×$P_{th}$ (making radiation pressure gain less than mechanical loss) so that the optical power only monitors the radial motion of the microtoroid induced by acoustic wave without significant interference with its dynamics. These sub-threshold measurements allow us to quantify the radial motion induced by the acoustic excitation ($\Delta R_{PZT}$) and its relation with injection locking independent of the specific actuator and configuration used to transfer the acoustic energy.

Fig. 3(c) shows the spectrum of the optical output power at eight different injection frequencies. Here $f_{PZT}$ is changed from $f_{OMO,2}$-1.36 kHz to $f_{OMO,2}$+1.40 kHz. As expected within a frequency range ($\Delta f_L$=2.76 kHz) around $f_{OMO,2}$ (known as lock range) the injection locked optomechanical oscillation frequency is equal to $f_{PZT}$ and follows its variations. Fig. 3(d) shows the frequency pulling effect at the edge of the lock range (again for mode-2). When $f_{PZT}$ is tuned slightly above and below the edge of the lock range, the oscillator is quasi-locked and the RF spectrum consists of a series of closely spaced decaying beat frequencies in the vicinity of $f_{OMO}$. This is a well-known effect that is studied in the context of electronic oscillators [20,21] and is also observed in optically injection locked OMOs [16]. We have carefully measured the lock-range for both mechanical modes injection locked to the PZT based on configurations shown in Fig. 2. Fig. 4(a) and (b) show the measured lock-range as a function of $P_{PZT}$ for mode-1 and mode-2, respectively. The relation between RF power and injection strength for each configuration is complicated and requires a full 3D FEM analysis of the whole system (PZT+silicon chip+microtoroid+glue tape). However the overall variation of the lock range for different modes and configurations can be explained based on the amplitude and the direction of mechanical vibrations generated by the PZT. For the selected PZT RF voltage modulates its thickness mode, configuration-1 and 2 generate mechanical vibrations along z-axis. As such for these configurations locking mode-2 requires more RF power because the vibrations along z-axis couple much more efficiently to mode-1 compare to mode-2. Also for both modes configuration-1 provides stronger injection compared to configuration-2 due to larger distance between PZT and OMO in configuration-2. Configuration-3 (where PZT is 90 degree rotated compared to configuration-2) provides the weakest injection as the resulting mechanical displacements are perpendicular to the displacement associated with mode-1 and mode-2.

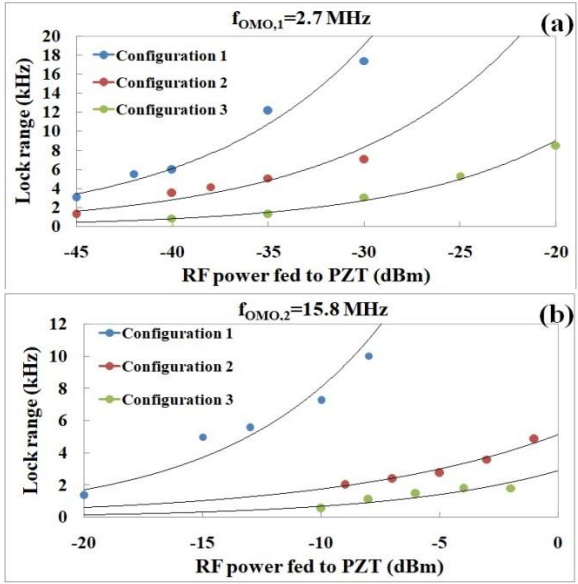

Fig. 4. Measured lock-range for mode-1 and mode-2 using configurations shown in Fig. 2 as a function of $P_{PZT}$.

Characterization of the lock range as a function of $P_{PZT}$ (RF power delivered to the PZT) for a given configuration is important and useful for designing injection locked OMOs. However in order to understand and evaluate the variation of lock range based on the general theory of injection locked oscillators (developed in the context of electronic oscillators), we need to characterize its behavior as a function of locking strength (as opposed to $P_{PZT}$).

In electronic oscillators and optically injection locked OMOs this is an easy task because the injected signal and the force that drives self-sustained oscillations are identical (voltage and optical power respectively). However when OMO is injection locked via acoustic waves, the injected signal is the RF power (or voltage) applied on the PZT while the driving force is the circulating optical power inside the cavity. Moreover the strength of the mechanical stimulation of the corresponding mode strongly depends on the PZT characteristics and the configuration used to transport the acoustic wave to OMO.

To characterize the lock range as a function of injection strength independent of excitation efficiency in a specific configuration, we used relative radial oscillation amplitude ratio defined as $\eta=\Delta R_{PZT}/\Delta R_{RP}$. Here $\Delta R_{RP}$ is the radial oscillation amplitude of the microtoroid driven by the radiation pressure ($P_{o,in}>P_{th}$) in the absence of external acoustic excitation ($P_{PZT}=0$). $\Delta R_{PZT}$ is the radial oscillation amplitude of the microtoroid induced by the acoustic wave (generated by the PZT) in the absence of self-sustained optomechanical oscillations ($P_{o,in}<P_{th}$). In other words $\Delta R_{PZT}$ and $\Delta R_{RP}$ are the radial oscillation amplitudes of the optical path length generated by acoustic energy and radiation pressure transferred to the corresponding mechanical mode respectively. As shown in Ref. 14 in general radial oscillation amplitude ($\Delta R$) of the optical path length is related to the measured optical modulation depth ($M$) through $\Delta R=(M\times D)/(2\times\Gamma\times Q_L)$ where $Q_L$ is the loaded quality factor of the optical mode, $M$ is the measured modulation depth and $\Gamma$ is the corresponding modulation transfer function that is ~1 when $f_{OMO}<<c/\lambda_0 Q_L$ (a condition valid for both modes studied here). Using the above mentioned relation we calculated $\Delta R_{RP}$ for each $P_{o,in}$ ($>P_{th}$) by measuring the modulated optical power when $P_{PZT}=0$. Then for each $P_{PZT}$ we calculated $\Delta R_{PZT}$ by measuring the modulated optical power while keeping $P_{o,in}$ below threshold to prevent radiation pressure driven oscillation (in this case optical power only served as a probe to monitor the microtoroid motion through optical modulation). Since $\Delta R_{PZT}$ is a measure of the actual acoustic energy transferred to the corresponding mechanical mode, behavior of the lock range as a function of $\eta$ is independent of efficiency of the PZT and acoustic energy transfer. Once the lock range is characterized as a function of $\eta$, finding the optimal configuration and actuation mechanism for minimizing the RF power required for achieving certain value of $\eta$ can be addressed separately using acoustic-mechanical design and optimization techniques.

Based on general theory of injection locking for self-sustained oscillators [16, 21], the lock range can be written as:

$$\Delta f_{lock} = \Delta f_{mech} \frac{\Delta R_{PZT}}{\Delta R_{RP}} \left[1-\left(\frac{\Delta R_{PZT}}{\Delta R_{RP}}\right)^2\right]^{-\frac{1}{2}} \quad (1)$$

Where the $\Delta f_{mech}$ ($\approx f_{OMO}/Q_{mech}$) is the intrinsic line-width of the passive mechanical resonator. Figure 5 shows the measured lock range as a function of $\eta$ for the two different mechanical modes using different acoustic excitation configurations. For the 1st mode, $f_{OMO,1}$=2.7 MHz and $\Delta f_{mech}$=23 kHz, for the 2nd mode, $f_{OMO,2}$=15.8 MHz and $\Delta f_{mech}$=44 kHz. The solid bands represent the theoretical prediction based on Eq. (1) and taking into account the error associated with measuring lock range and $\eta$. The standard deviation for the measured lock range was 0.328 kHz and 4.188 kHz for the first and second mode respectively. The uncertainty of $\eta$ is proportional to $[\Delta R_{PZT}\times\delta(\Delta R_{RP})-R_{RP}\times\delta(\Delta R_{PZT})]/(\Delta R_{RP})^2$. $\Delta R_{PZT}$ and $\Delta R_{RP}$ are calculated based on measured modulation depth and therefore detected modulated optical power ($P_{mod,RF}$); so $\delta(\Delta R_{RP})$ and $\delta(\Delta R_{PZT})$ are proportional to $\delta P_{mod,RF}$ that was about ±0.8 dBm for all measurements.

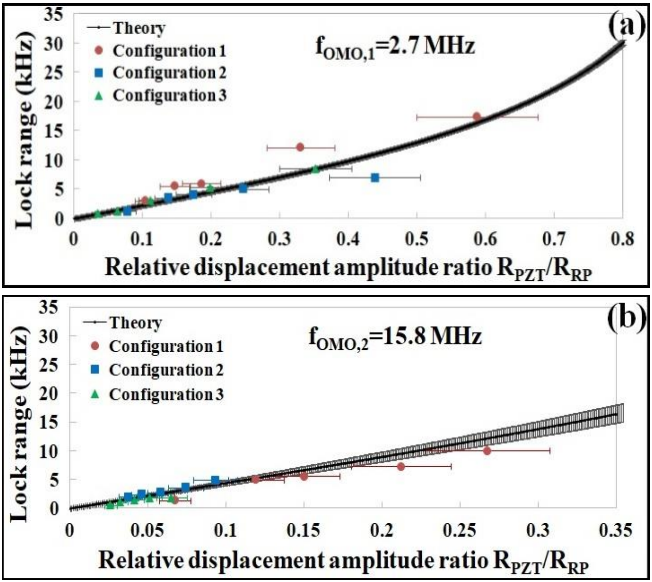

Fig. 5. Measured lock range plotted against the relative displacement ratio ($\eta$) based on different configurations for: (a) Mode-1 ($f_{OMO,1}$=2.7 MHz) and (b) Mode-2 ($f_{OMO,2}$=15.8 MHz). The solid lines are theoretical estimation based on Eq. (1).

Fig. 6 shows the measured spectrum of the OMO optical output power as a function of $P_{PZT}$ and $\eta$ for mode 1 and using configuration-2 when $f_{PZT}$-$f_{OMO}$≈1.35 kHz. Above -65dBm ($\eta$>0.011) injection pulling begins and at ~-45 dBm ($\eta$=0.111) the OMO is locked to the PZT.

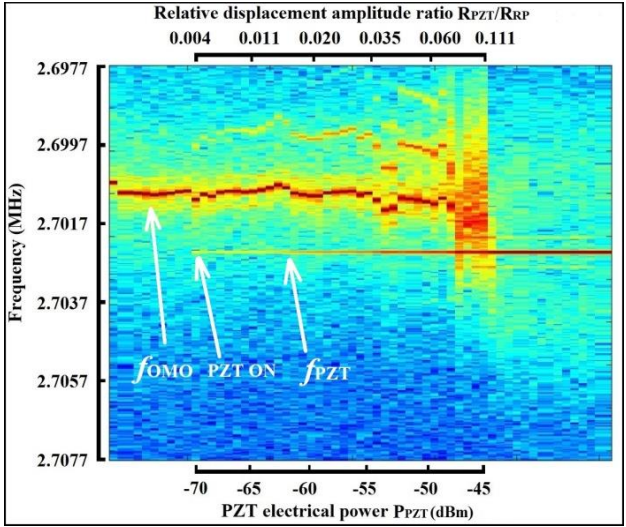

Fig. 6. RF spectrum of the OMO plotted against $P_{PZT}$ and the relative displacement ratio $\eta$ for the 1st mode and using configuration-2.

Fig. 7(a) and (c) show the temporal behavior of the measured phase difference between the RF signal driving the PZT and the OMO output ($\Delta\Phi=\Phi_{OMO}-\Phi_{RF}$) in the presence (ON) and absence (OFF) of the injection signal when $f_{PZT}=f_{OMO}$. While based on basic injection locking theory [20,21] the phase difference between OMO and the injected signal should be zero when $f_{PZT}=f_{OMO}$ (assuming $P_{PZT}$ is large enough to lock the OMO), here $\Delta\Phi$ is -100° and 90° at $f_{PZT}=f_{OMO}$. This phase off-set is caused by the delay between the injection signal (the acoustic excitation fed to the toroid) and the RF drive due to the response time of the PZT (RC time constant) and acoustic wave propagation from PZT to the toroid. Fig. 7 (b) and (d) show the variation of $\Delta\Phi$ as a function of frequency detuning ($\Delta f=f_{OMO}-f_{PZT}$) within 5.54 kHz and 2 kHz lock range.

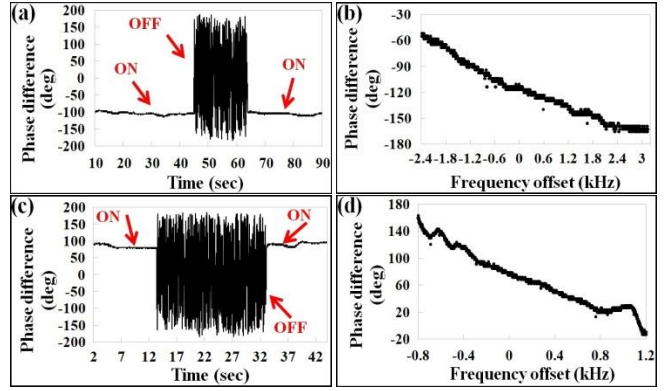

Fig. 7. *Left column*: temporal behavior of the measured phase difference between RF signal fed to PZT and the OMO optical output power ($\Delta\Phi=\Phi_{OMO}-\Phi_{RF}$) in the presence (ON) and absence (OFF) of injection signal when $f_{PZT}=f_{OMO}$ for (a) the 1st mode measured using configuration-1 and (c) the 2nd mode measured using configuration-2. *Right column*: Measured $\Delta\Phi$ plotted against $f_{OMO}-f_{PZT}$ for (b) the 1st mode measured using configuration-1 and (d) the 2nd mode measured using configuration-2. Here $P_{PZT}$ = -42 dBm and $\eta$=0.147 for (a) and (b), and $P_{PZT}$ = -9 dBm and $\eta$=0.037 for (c) and (d).

## 4. THEORETICAL MODELING OF LOCK RANGE AND PHASE VARIATIONS

We have used the time-domain classical theory of optomechanical oscillation [2] to calculate the lock range as a function of the $P_{PZT}$. The optomechanical oscillation can be described by two coupled differential equations that govern the temporal variation of radial component of the microtoroid displacement and the circulating (resonant) optical power. These two equations are coupled through radiation pressure of the circulating optical power that acts as a radial force on the microtoroid and is controlled by the optical frequency detuning ($\Delta\omega_0=\omega_{laser}-\omega_0$, where $\omega_0$ is the resonant frequency of the selected optical mode). The presence of an acoustic excitation is equivalent to an additional harmonic external force ($F_A(t)=F_{A0}\cos(\Omega_{PZT}t)$) that is added to the optical force (radiation pressure). The resulting coupled differential equations can be written as:

$$m_{eff}\ddot{r}+b\dot{r}+kr=\frac{2\pi|A(t)|^2 n}{c}+F_{A0}\cos(\Omega_{PZT}t) \quad (2)$$

$$\dot{A}(t)+A\left\{\frac{\alpha c}{n}-i\left[\Delta\omega_0+\frac{\omega_0 r(t)}{R}\right]\right\}=iB\sqrt{\frac{\alpha c}{n\tau_0}} \quad (3)$$

Here $m_{eff}$ is the effective mass associated with the radial component of the corresponding mechanical mode, $r(t)$ is the radial displacement of the microtoroid, $b$ is the mechanical dissipation and can be inferred from the measured sub-threshold acoustic bandwidth, $k$ is the spring constant, $|A(t)|^2$ is the circulating optical power, $n$ is the refractive index of silica at 1550 nm wavelength, $\alpha$ is the optical loss in the cavity, $R$ is the radius of the optical path (~radius of the mocrotoroid), $B$ is the input pump field (normalized such that $|B|^2$ is the optical pump power). $F_{A0}$ is the amplitude of the equivalent radial force corresponding to the acoustic excitation. In order to calculate the optomechanical oscillation frequency as a function of the RF power that drives the PZT ($P_{PZT}$) we have calculated the relation between $F_{A0}$ and $P_{PZT}$ using finite element modeling (see supplementary information). Since modeling configuration-2 and-3 requires a relatively large model and therefore long simulation time. We have limited our calculation to mode-1 excited via configuration-1. The cylindrical symmetry of configuration-1 allows reducing the simulated zone without significant impact on the outcome. The amplitude of the radial force inserted on the microtoroid when the PZT is driven at $\Omega_{PZT}=2\pi f_{OMO,1}$ can be written as (see supplementary information):

$$F_{A0} = 3.3\times10^{-9} \times 10^{\frac{P_{PZT}}{20}} \quad (4)$$

Here $P_{PZT}$ is in dBm and it has been assumed that impedance of the RF source and the PZT are 50 and ~32 ohms respectively (based on the actual PZT and signal generator used in our experiment). Fig. 8 shows the calculated lock range as a function of $P_{PZT}$ using equation 2 and 3. The red dots are experimental results for the 1st mode and configuration-1 (extracted from Fig. 4). The good agreement between experimental and calculated results shows the validity of our assumptions and therefore the usefulness of this simple model for predicting the locking behavior of the optomechanical systems.

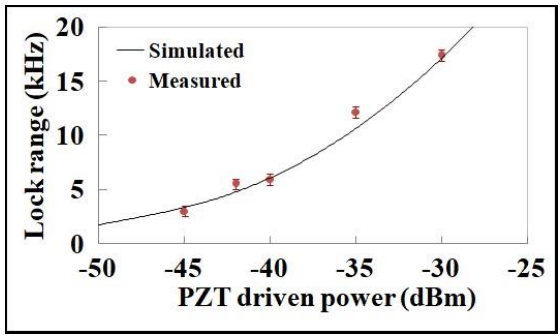

Fig. 8. Calculated (black solid line) and measured (red circles) lock-range for mode-1 injection-locked using configuration-1. Here, $b=0.92\times10^{-6}$ kg/s, $n=1.46$, pump laser frequency is fixed 0.38 FWHM larger than the resonant frequency of the toroid, so $\Delta\omega_0=0.38\omega_0/Q_{L,1}$, $\omega_0$ can be inferred from $\lambda_{01}$.

Fig. 9 shows the calculated the phase difference between $F_A(t)$ and $r(t)$ (=$r_0\cos(\Omega_{PZT}t+\gamma)$) for mode-1 and configuration-1 using the same equations, here $P_{PZT}$=-42 dBm, and all parameters are same as that used in the experiment to obtain fig. 7 (a) and (b). Note that in the experiment we measured the phase difference between $r(t)$ and $V_{RF}(t)$(=$V_{RF,0}\cos(\Omega_{PZT}t-\theta)$), so although the behavior of the simulated (Fig. 9) and measured (Fig. 7(b)) results are in good agreement, as opposed to the measured results the simulated phase off-set is zero ($\gamma$=0 at $f_{PZT}=f_{OMO,1}$). While the relation between amplitude of $F_A(t)$ and $V_{RF}(t)$ can be estimated using finite element modeling, their phase difference ($\theta$) involves more advanced modeling tools and computational resources. So by using $\gamma$ instead of $\gamma+\theta$ in our simulation we are ignoring the delay between the RF voltage applied on the PZT and the acoustic excitation on the microtoroid (we assume $\theta$=0).

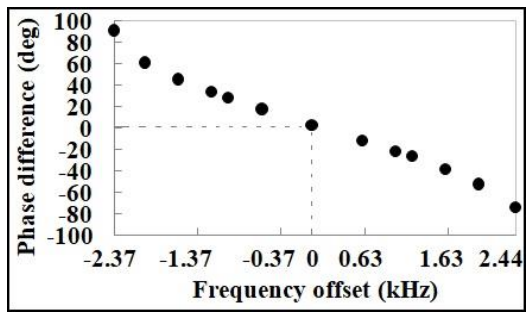

Fig. 9. Calculated phase difference between the $F_A(t)$ and $r(t)$ for mode-1 and configuration-1 using equation 2-3, here $P_{PZT}$=-42 dBm.

## 5. CONCLUSION

For the first time we have demonstrated injection locking of a radiation-pressure driven oscillator to an electromechanical transducer via acoustic waves. Even without acoustic impedance matching and optimizing the energy transfer between the transducer and the OMO, a lock range of 17 kHz has been achieved with only 1 microwatt (-30 dBm) RF power. We expect the required RF power for a carefully designed impedance matched system to be significantly lower. For example by eliminating the acoustic reflection between PZT and the silicon chip in configuration-1 (using an impedance matching layer), mode-1 can be locked within 17 kHz range with an RF power as low as 270 nanowatt. Note that even in the absence of acoustic loss only a small portion of the transmitted acoustic energy is absorbed by the OMO (due to the small interaction cross-section of the OMO). As such with the same level of RF power multiple OMOs on a chip can be locked to a single RF source. Moreover employing on-chip electromechanical transducers based on piezo electric thin films and interdigital electrodes, enables excitation of various types of surface waves that may transfer the acoustic energy to the OMO more efficiently. Additionally integrated acoustic waveguides and photonic crystals can be used to improve the directivity of the acoustic energy transferred to the target OMO. Using this approach the acoustic energy from one transducer can be distributed among several OMOs or multiple transducers can be independently locked to groups of OMOs.

These possibilities combined with the fact that injection locking via acoustic waves does not require power hungry optical modulators and direct physical contact with the OMO, makes this approach superior to the previously demonstrated techniques (based on optical modulation and direct application of electrostatic force) in particular for locking nano scale OMOs and synchronization of OMO networks. While the physics and behavior of a network of synchronized OMOs has yet to be explored, theoretical studies on network of synchronized oscillators has revealed very interesting properties that are promising for communication and sensing applications. For example it has been shown that the frequency precision of a network of regenerative oscillators perturbed by $N$ independent noise sources is improved by a factor of $N$ [27].

While we did not measure the phase-noise of the locked OMO (due to lack of access to a phase-noise analyzer), based on previous results (injection locking both based on optical modulation [16] and direct electrostatic force [25]), it is clear that in addition to synchronization and frequency control, injection locking via acoustic wave can reduce the phase noise of the OMO proportional to the power and phase noise of the RF source that generate the acoustic waves.

This work was supported by National Science Foundation (grant number 1055959)

**Acknowledgment**. We thank Mr. Behsan Behzadi for his help with the calculations and modeling.

# Supplementary material for "Injection locking of optomechanical oscillators via acoustic waves".

## Relation between RF driving power and equivalent acoustic force

As mentioned in section 4, in order to simulate the variation of lock range and relative oscillation phase as a function of the RF power fed to the PZT ($P_{PZT}$) using coupled differential equations 2 and 3, the radial component of the equivalent acoustic force ($F_{A0}$) experienced by the microtoroid should be known as a function of $P_{PZT}$. Here we present the calculation procedure that leads to equation 4. We used finite element modeling (COMSOL-mechanical package) for these calculations. Since modeling configuration-2 and -3 requires a relatively large model and therefore long simulation time, we have limited our calculation to mode-1 ($f_{OMO,1}$=2.7 MHz) excited via configuration-1. The cylindrical symmetry of configuration-1 allows reducing the simulated zone without significant impact on the outcome. Fig. S1 shows the configuration used in the simulation where the silicon and acrylic tape thicknesses are selected based on experimental values while the area of the chip below OMO is reduced to 80×80 μm. The thickness of the PZT is also reduced as the software allows adjusting the PZT parameters such that its response is similar to that of the actual PZT without the need to model the whole PZT thickness.

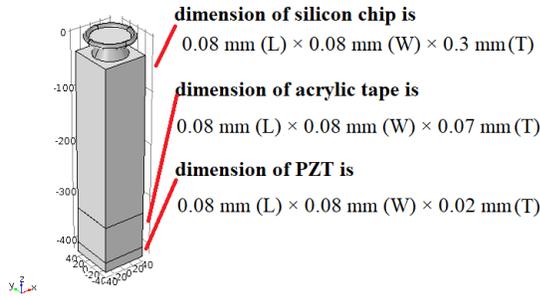

Fig. S1. The model used for calculating the relation between $F_{A0}$ and $P_{PZT}$ using finite element modeling.

As the software did not allow direct calculation of the amplitude of the force experienced by a certain mechanical mode ($F_{A0}$) we used amplitude of the radial displacement of toroidal section to find the relation between $F_{A0}$ and $P_{PZT}$. First we calculated the relation between an external harmonically varying radial force ($F_A = F_{A0}\cos(\Omega_{PZT})$ where $\Omega_{PZT}=2\pi f_{OMO,1}$) inserted on the toroidal section of the microresonator and the resulting displacement amplitude ($d_r = d_{r0}\cos(\Omega_{PZT})$). Fig. S2 shows that $d_{r0}$ varies linearly with $F_{A0}$ with a slope of 3 mm/N.

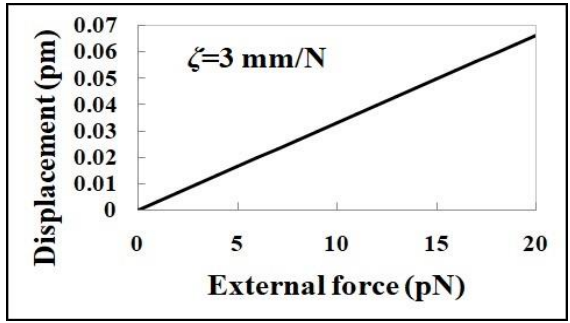

Fig. S2. Calculated displacement amplitude ($d_{r0}$) as a function of the amplitude of the external force ($F_{A0}$) inserted on the toroidal section of the microresonator.

Next we calculated the $d_{r0}$ as a function of the amplitude of the RF voltage (peak voltage) applied on the PZT ($V_{RF}=V_{RF,0}\cos(\Omega_{PZT}t-\theta)$). The PZT parameters and mechanical boundary conditions were selected such that the acoustic waves generated by the PZT for a given RF power is the same as the actual PZT. This was done by exciting the same mechanical mode (thickness mode) of the PZT that is excited in the experiment, and adjusting the piezoelectric and mechanical properties of the PZT according to its actual specifications (extracted from the spec sheet). Fig. S3 shows that $d_{r0}$ varies linearly as a function of PZT driving voltage with a slope of 39 pm/V. As mentioned earlier the limitation of our software did not allow extracting the phase difference ($\theta$) between $V_{RF}$ and $d_r$.

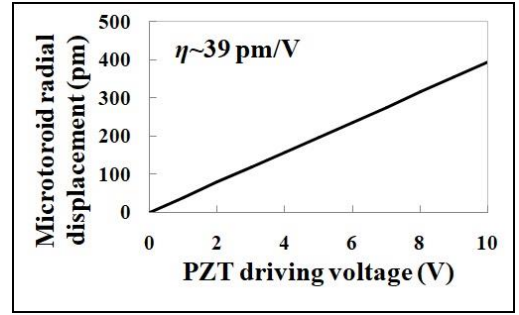

Fig. S3. Calculated displacement amplitude ($d_{r0}$) as a function of the amplitude of the voltage ($V_{RF,0}$) applied on the PZT.

As such we concluded that the amplitude of the equivalent radial force is related to the amplitude of the applied RF voltage via:

$$F_{A0} = \frac{\eta}{\zeta} \times V_{RF,0} = \frac{39 \times 10^{-12}}{3 \times 10^{-3}} \times V_{RF,0} = 13 \times 10^{-9} \times V_{RF,0} \quad (S1)$$

The impedance of the RF source is 50 ohms and the estimated impedance of the PZT at 2.7 MHz (based on its value at resonance) is about 32 ohms. So $V_{RF,0}$ in above equation can be replaced by $P_{PZT}$:

$$F_{A0} = \left(\frac{\eta}{\zeta}\right)\sqrt{\frac{2Z}{1000} \times 10^{(P_{PZT}/10)}} = 3.3 \times 10^{-9} \times 10^{(P_{PZT}/20)} \quad (S2)$$

Here the unit of $P_{PZT}$ is dBm, $F_{A0}$ is in Newton.